# AN ACTIVE CHAOTIC MICROMIXER INTEGRATING THERMAL ACTUATION ASSOCIATING PDMS AND SILICON MICROTECHNOLOGY

*Olivier Français[(1)], Marie Caroline Jullien[(2)], Lionel Rousseau[(3)], Patrick Poulichet[(3)], Serge Desportes[(4)],Assia Chouai[(4)], Jean Pierre Lefevre[(4)], Jacques Delaire[(4)]*

(1) CNAM-ESCPI, JE2405, 292 Rue St Martin, 75141 Paris Cedex 03, France
(2) SATIE, ENS de Cachan, 61 Av du Prdt Wilson, 94235 Cachan, France
(3) ESIEE, 2 Bd Blaise Pascal, 93162 Noisy le grand, France
(4) PPSM, ENS de Cachan, 61 Av du Prdt Wilson, 94235 Cachan, France

## ABSTRACT

Due to scaling laws, in microfluidic, flows are laminar. Consequently, mixing between two liquids is mainly obtained by natural diffusion which may take a long time or equivalently requires centimetre length channels. To reduce time and length for mixing, it is possible to generate chaotic-like flows either by modifying the channel geometry or by creating an external perturbation of the flow.

In this paper, an active micromixer is presented consisting on thermal actuation with heating resistors. In order to disturb the liquid flow, an oscillating transverse flow is generated by heating the liquid. Depending on the value of boiling point, either bubble expansion or volumetric dilation controlled the transverse flow amplitude.

A chaotic like mixing is then induced under particular conditions depending on volume expansion, liquid velocity, frequency of actuation… This solution presents the advantage to achieve mixing in a very short time (1s) and along a short channel distance (channel width). It can also be integrated in a more complex device due to actuator integration with microfluidics.

## 1. INTRODUCTION

With the help of microtechnology applied to microfluidics, miniaturized chemical and biological analysis units offer a wide range of opportunities in term of applications and performances [1]. Due to the scaling laws, flow behaviour in micro scale is drastically laminar and is associated with very low Reynolds number (<1). The Reynolds number (Re) is the ratio between inertial forces compared to viscous forces in a fluid:

$$R_e = \frac{\rho V L}{\mu}$$

with ρ the fluid density, V the mean velocity, L characteristic channel dimension and μ the fluid viscosity.

In consequence, in small channels, the familiar turbulent mixing of macroscopic devices does not occur. In case of two fluids flowing in a channel side by side, a well defined interface appears and only diffusion creates mixing. In microfluidic, diffusion time can be drastically reduce by acting on interface fluids.

Several micromixers have been studied and published so far. Passive ones are based on asymmetric geometric disturbance to induce circular flow between two fluids [2] [3], or for example on the division of the liquids into tiny drops which are collected, after that in each drop, the liquids mix (due to internal friction) [4]. By comparison, active micromixers present better performance in term of mixing efficiency, time and length of mixing. To achieve mixing, several kinds of actuators exist and mainly stretch liquids. Disturbance in flow can be generated in several manners [5] : by ultrasound [6], vibration, electrowetting, magneto-hydrodynamic action or external pressure.

Here, an original actuator is presented. It is based on thermal actuation by dilation effects. It offers the capability to integrate the actuator with the channel and keep an actual control on the mixer behaviour. By creating interferences in the liquids interface, chaotic phenomenon occurs and induces exponential growth of the contact interface [7] [8] [9] .

This paper will present principle description of the mixer and technology associated. It also presents first results of the mixing efficiency as a function of thermal actuator characteristics.

## 2. STRUCTURE AND WORKING PRINCIPLE

The micromixer is based on a classical microfluidics channel obtained by PDMS moulding and bounded to a silicon or glass wafer [10]. It consists of a main channel where two fluids flow side by side. Due to diffusion time, a channel length of several centimeters is needed to obtain a complete diffusion between the two fluids. To reduce this length, a thermal actuator is added to increase the interface of the two fluids by stretching and folding it alternatively.

Thus, in order to disturb the interface between the two fluids, an intersection is achieved with a secondary channel perpendicular to the main one. By this way, an oscillating transverse flow can be generated and may modify the position of the liquids interface [7]. Depending on the amplitude and frequency of this transverse flow compared with the one in the main channel, an increase of the contact surface between the two fluids can be obtained. With low transverse flow amplitude, no mixing is obtained. The effective mixing occurs when disturbance shape is sheared by the main flow. Thus, transverse flow may have amplitude greater than the channel width and velocity at least greater than the two fluids. Mixing is achieved by this way in short length similar to channel width and can be dynamically controlled.

To activate the transverse flow, an original thermal actuator has been integrated in the chip. On each side of the transverse channel, a circular microcavity is placed and associated to a heating resistance. Thus, heating the fluid will generate a volume expansion in the cavity and induce a fluid displacement in the transverse channel. By controlling the frequency and voltage value applied on the heating resistance, the oscillating transverse flow is generated. At microscales, thermal diffusion is a fast process, leading to a quasi simultaneous response of the system with respect to the actuation. In fact, with silicon substrate, a cut frequency of about 10Hz had been obtained for a thermal cavity surface of around 3.1mm² for 100μm thick. Compare to macroscale, this cut frequency value is very interesting.

This thermal actuator is obtained by deposition of a conductor on the substrate. It is in contact with the cavity filled by a liquid and connected to the microfluidic channel. Two thermal cavities are needed (each side) and activated one after the other (Fig. 1).

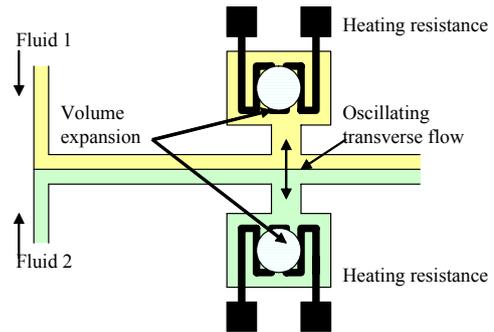

**Figure 1 : Principle of the micromixer**

Thermal properties of the heated fluid will modify the mixer behaviour. Here, two cases have been considered. The first one supposed that a temperature above the boiling point is reached. To achieve the mixing, the cavity temperature was always above the boiling point with an oscillating value. Thus, the transverse flow was generated by a bubble volume variation. This case has been used with water, by heating this liquid above 100°C, a gas bubble was created and caused transverse flow across the main channel. This principle gave very large transverse flow due to bubble volumetric expansion coefficient. But, it was difficult to control. In figure 2, bubbles located in transverse channel are visible. One bubble is growing by activating the heating resistance; the other is collapsing due to passive cooling across the device.

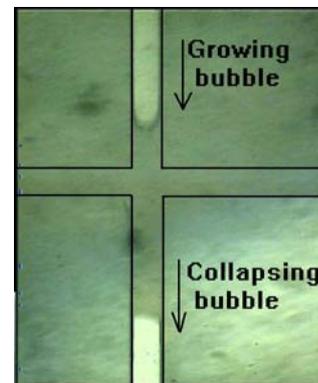

**Figure 2 : Photography of bubbles expansion and collapsing with water**

The second one supposed that cavity temperature was always under boiling point. In this case, no bubble was generated, but only fluid volumetric dilation activated the transverse flow. A mixture of 70% of Glycerol with 30% of DI water has been used. The advantage is a very robust control of the amplitude generated, and here, associated to fluorescein in order to label one of the two liquids, characterisation of the mixing was simplest due to low value of the diffusion coefficient (fig. 3).

The fluid volumetric dilation $\Delta V_{ol}$ in heating cavity is then concentrated in the channel associated. By applying the volume expansion value in the channel, the average length transverse flow $\langle x \rangle$ can be estimated:

$$\Delta V_{ol} = \alpha.\Delta T.\left(\frac{\pi d^4.h}{4}\right) = (h.h)^2.\langle x \rangle$$

With $\alpha$ the fluid dilation coefficient, $\Delta T$ the temperature increase, d the cavity diameter, h thickness of all microfluidic parts. The channel section is considered square.

A value of $0,425 \times 10^{-3}$ for $\alpha$ has been estimated (mixture between glycerol and water value), h is equal to 100μm and d is 2mm. By this way, the average length $\langle x \rangle$ is connected to $\Delta T$ with the relation :

$$\langle x \rangle = 13.\Delta T \quad (\mu m)$$

In case of laminar flow, the transverse flow length will be the double of $\langle x \rangle$. Thus, a $10^{\circ C}$ temperature increase will generate a transverse flow of around 260μm of maximum amplitude. This value is similar to measurement achieved but is slightly over-estimated due to temperature gradient in heating cavity. Numerical simulation are to be made to precise modeling.

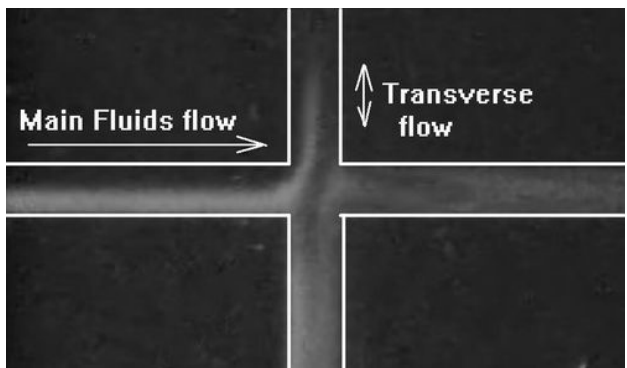

**Figure 3 : Photography of transverse flow with glycerol**

## 3. FABRICATION PROCESS AND ASSOCIATED ELECTRONICS

The thermal micromixer is composed of two parts. The electrical part is made on silicon by metal deposition and the micro fluidic part is made in PDMS [10]. By a classical method, the two parts are sealed together to avoid any leak.

First Silicon wafers were placed at 1050°c during 40 minutes on a steam furnace to obtain an oxide layer ($SiO_2$) of 0.4 μm thickness. This layer was used as an insulator for electrical parts. The wafers were then placed on RF magnetron sputtering and covered by 100 nm of chromium and 500 nm of gold. Chromium layer was used to allow a good adhesion on the substrate. A classical photolitography process was done to pattern the design of the heating resistances. The gold layer and chromium layer were etched successively by wet etching to obtain the heating resistance.

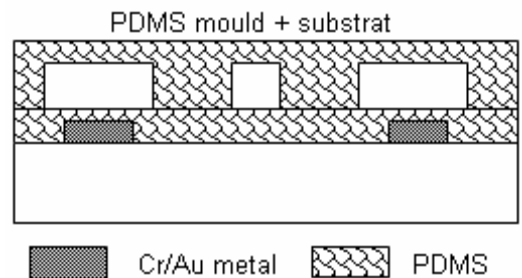

**Figure 4: Cut view of the process microchip**

For the microfluidic part, polydimethysiloxane (PDMS) was used. This product offers a large number of possibilities for achieving a prototype. First a mould had to be done. It was realised by using Deep Reactive Ion Etching (DRIE) technique. The DRIE technique [11] allows etching silicon with high aspect ratio and gives vertical side walls with better aspect than polymer technique. Concerning PDMS [12], it is composed of two products: base and reticulant. They were mixed together in proportion of 10:1 v/v. Before pouring PDMS on the mould, all the surface of the mould was treated with dimethysilixone vapour deposition during 30 minutes to make easier peeling off. The PDMS was poured in the mould and cured during two hours at 70 °C.

Then, all the PDMS micro-fluidics parts were delicately peeled off and sealed on the silicon wafer with hot resistance. To insulate the heating resistances from the liquid, a thin layer of PDMS is spin coated (fig. 4). This wafer with PDMS was cured in an oven at 70°C during 2 hours . To perform a permanent bonding between silicon wafer with heating resistances and PDMS microfluidics, activation by a plasma cleaner ($O_{2 \, gas}$) was done on both PDMS surfaces during 30 seconds at a pressure of 300 mtorr. Plasma-treated samples were immediately bonded together with activated surfaces facing each other and then baked in an oven at 70 °C for 30 minutes (Fig. 5).

An homemade electronic circuit is used for generating the voltages applied on the two heating chambers. This circuit enables to choose frequency, duty cycle and voltage. These parameters are adjusted in order to find the best conditions leading to the best mixing. Therefore, the power applied on the chamber can be controlled. The two heating resistances are in fact used in an opposite mode. An adjustable delay has been added to eventually use dissymmetric voltage. The frequency can be varied from 0.1 Hz to 100 Hz and the duty cycle is adjusted from 0.01 to 0.99. The heating resistor value is about 88 Ω. Thus, with a 30 V source and a duty cycle of 0.5, an electrical power of 6 W can be reached.

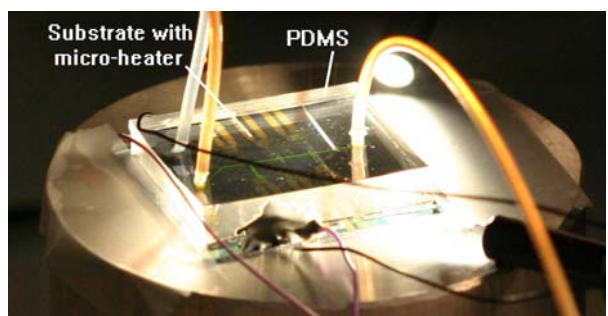

**Figure 5 : Micro-chip with microfluidic and electronic connection**

## 4. MEASUREMENT AND FIRST RESULTS

The performance of the mixing depends on the liquids velocity, length of transverse disturbances and frequency associated. Presently, the use of thermal actuators induces a limitation in frequency response because of substrate thermal resistivity. Consequently, two kinds of substrates have been used : one is a glass wafer which leads to a low frequency response (around 1Hz) but enables back side characterization, the other one is a silicon substrate which leads to a better frequency response (close to 10Hz).

Once the linear dependence of temperature versus injected electrical power had been validated [13], a design was developed to obtain the thermally activated micromixer. All channels have a section of 100 μm*100 μm. Length of the main channel and transverse channel was 4 mm. The two thermal cavities were circulars with a diameter of 2 mm and 100 μm thick. The heating resistance value was 88Ω.

In order to characterize the micromixer, specific liquid with low diffusion value has been used. So Glycerol in 70% concentration associated with 30% DI water (fluorescein is added for the second fluid) has been injected in the device (Fig. 3). The fluid flow was constant and fixed at 10 μl/hr in the main channel corresponding to an average velocity of 0.28 mm/sec. The shape voltage applied on each resistance was square with a duty cycle of 0.5. Heating resistance were activated in opposite way. Study was done for a frequency from 1 Hz to 7 Hz with a voltage value varied from 0V to 30V corresponding to a total power varying from 0 to 5.1 W.

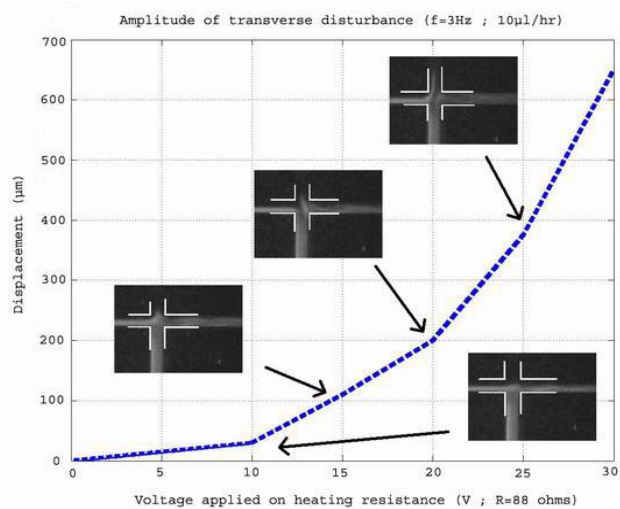

**Figure 6 : Transverse flow amplitude evolution with applied voltage for a main fluids flow of 10μl/hr**

First measurements have been achieved in order to validate the actuator principle. Depending on the frequency and voltage values, amplitude of the transverse disturbances was modified. For a fixed frequency, this amplitude increases quadratically with the applied voltage (fig. 6). Furthermore, mixing occurred for disturbance amplitude greater than the channel size. Figure 6 shows the evolution of the transversal displacement as a function of the applied voltage, for a given frequency (3Hz). The pattern at crossing is also shown for different couple (displacement, applied voltage). The evolution is not linear, however, as expected, the trend is an increase of the displacement versus an increase of the applied voltage. Note that this plot has to be handle with care. Indeed, the frequency plays a key role on the transversal displacement, i.e. the interface can be stretched and folded several time at crossing, this is an explanation for a non linear increase of the displacement as a function of applied voltage, even at a given frequency. Consequently, different couples (frequency, applied voltage) may lead to identical displacements. This will be further represented on figure 7.

The principle of the mixer has been validated, the efficiency of mixing was quantitatively measured using classical image treatment. At the exit of the crossing, when the main flow has been perturbed by the transversal oscillating flow, images were recorded and further processed. For each image, the histogram of concentration distribution was calculated.

O. Français, C. Jullien, L.Rousseau, P. Poulichet, S. Desportes, A. Chouai, JP Lefevre, J. Delaire
*An active chaotic micromixer integrating thermal actuation associating PDMS and Silicon microtec.*

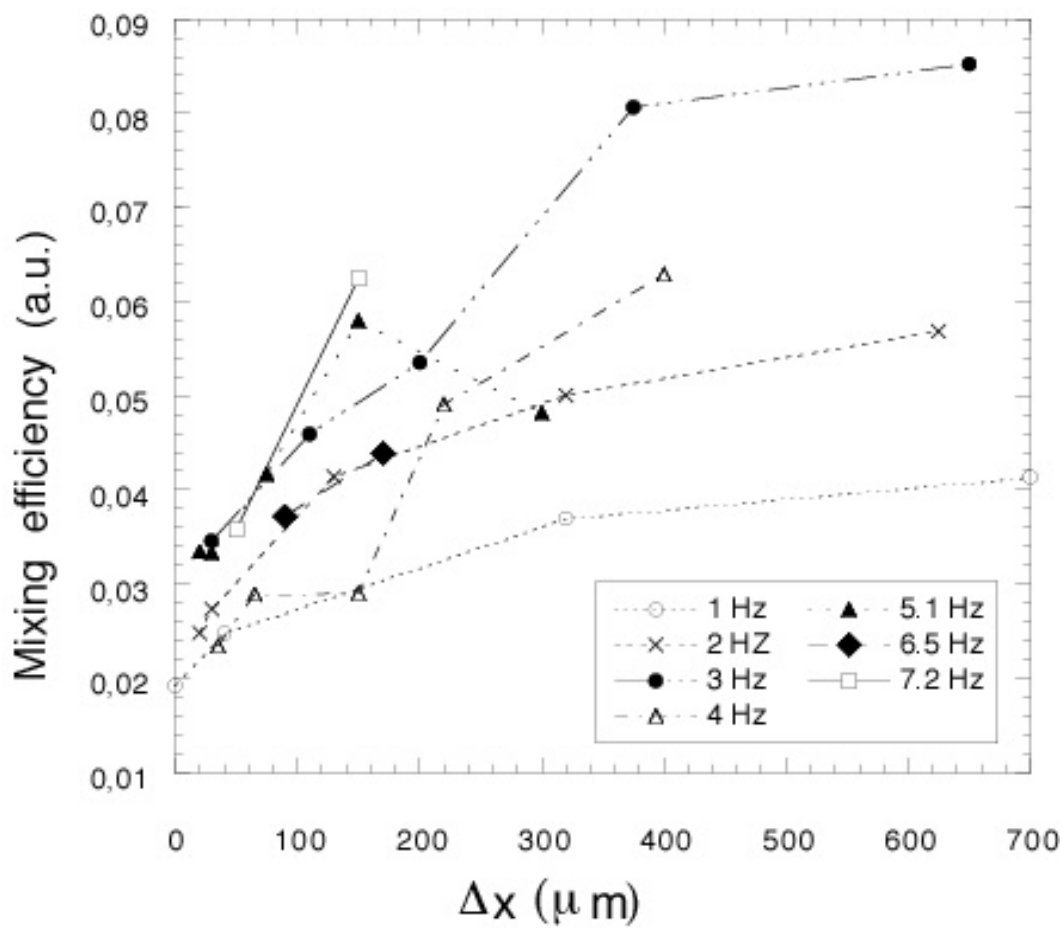

**Figure 7 : Mixing efficiency as a function of transversal displacement for different forcing frequencies.**

In a configuration in which there was no transverse oscillating flow, this histogram had a two-peaks shape, one peak corresponds to a high concentration of the dye, while for a full mixing of the two liquids, this histogram tends to a Gaussian. The root mean square of the distribution is a measure of mixing efficiency. Indeed, for poor mixing, the root mean square is larger than for a good mixing situation. In order to characterize the efficiency of the mixer as a function of both control parameters (frequency, electrical power), the reverse of root mean square is evaluated with transversal displacement (Fig. 7).

From figure 7, the relationship between mixing efficiency as a function of transversal displacement is represented for different forcing frequencies. The general trend, as expected, is an increase of mixing efficiency with respect to an increase of transversal displacement. Under the experimental conditions, i.e. a flow velocity of 0.28 mm/sec, the maximal mixing efficiency is achieved for a frequency of 3 Hz. The fact that for a given transversal displacement the mixing efficiency is not a constant underlines the strong influence of the frequency.

Indeed, the flow velocity and the forcing frequency are intrinsically connected, this may lead to a single or several stretching and folding of the two liquids interface inside the crossing, i.e. even if the displacement reaches a given value the interface can undergo several times the displacement.

The microfluidic device will be next used in quantitative detection of cations dedicated to potassium by fluorometric method.

### 5. CONCLUSION

In this paper, an active micromixer is depicted. Based on a chaotic like mixer, it used an integrated heating resistance to be activated. The thermal dilation obtained generates the transverse flow necessary to obtain the stretching and folding of the main flow.

Concerning manufacture, the microchip is realised using classical MEMS technology associated to PDMS moulding. By this way, the thermal actuator has been integrated with all microfluidic.

The phenomenon used to activate the micromixer has been described and shown to be dependant of fluid thermal properties. Two cases have been considered concerning the obtaining of the transverse with the thermal actuator. The first is based on boiling of the liquid and, thus, on bubbles expansion or collapse. The second used only liquid dilation with temperature without boiling. Experiments have validated the two principles.

To characterize the mixer, image treatment is used by calculating the histogram of concentration distribution. From first measurement, a good mixing region is shown related to actuator characteristics.